\documentclass[10pt,a4papper]{article}
\usepackage{indentfirst}
\usepackage[]{graphicx}
\usepackage{geometry} 
\begin{document}

\title{\textbf{The processes $e^{+}e^{-} \rightarrow K^{\pm} (K^{*\mp}(892), K^{*\mp}(1410))$ and $e^{+}e^{-} \rightarrow (\eta, \eta^{'}(958)) (\phi(1020), \phi(1680))$ in the extended Nambu-Jona-Lasinio model}}
\author{M. K. Volkov\footnote{volkov@theor.jinr.ru}, A. A. Pivovarov\footnote{tex$\_$k@mail.ru}\\
\small
\emph{Bogoliubov Laboratory of Theoretical Physics, Joint Institute for Nuclear Research, Dubna, 141980, Russia}}
\maketitle
\small

\begin{abstract}
The processes $e^{+}e^{-} \rightarrow K^{\pm} K^{*\mp}(892)$ and $e^{+}e^{-} \rightarrow \eta \phi(1020)$ are calculated in the framework of the extended Nambu-Jona-Lasinio model. The intermediate vector mesons $\rho(770), \omega(782), \phi(1020)$ and their first radially-excited states are taken into account. The obtained results are in satisfactory agreement with the experimental data. The predictions for the cross sections of the reactions $e^{+}e^{-} \rightarrow K^{\pm} K^{*\mp}(1410)$, $e^{+}e^{-} \rightarrow \eta^{'}(958) \phi(1020)$ and $e^{+}e^{-} \rightarrow \eta \phi(1680)$ were made.
\end{abstract}
\large
\section{Introduction}
In recent years, the processes of meson production at colliding of electron-positron beams at low energies ($\sqrt{s} \le 2$ GeV)
is actively investigated at various experimental facilities, for instance, BaBar, Belle, CMD-3 and others \cite{Aubert:2007ym, Ivanov:2016blh, Volkov:2016umo}.
Concerning theoretical research, the main
difficulty is inability to use the perturbation theory of quantum chromodynamics at low energies. Thus, one has to apply various
phenomenological models. They are generally based on the chiral symmetry of strong interactions and principles
of the vector meson dominance \cite{Volkov:2016umo}.

One of the most successful phenomenological models of this type is the Nambu-Jona-Lasinio (NJL) model
\cite{Ebert:1982pk, Volkov:1986zb, Ebert:1985kz, Klevansky:1992qe, Volkov:1993jw, Ebert:1994mf, Volkov:2006vq} and its new version called
the extended NJL model \cite{Volkov:2006vq, Volkov:1996br, Volkov:1996fk, Volkov:1999yi}, which allow one to describe mesons in
both ground and first radially-excited states. Unlike other phenomenological models they are based on the principle of
spontaneous breaking of chiral symmetry and include a minimal amount of arbitrary parameters.
This increases their prediction power, as shown in \cite{Volkov:2016umo}.

Until now, in the framework of the NJL models, the processes of non-strange meson production in electron-positron
annihilation have been described. In the present paper, it is shown that using these models one can successfully describe the meson production involving
s-quarks. The Feynman diagrams of the reactions considered in this work have an anomal quark triangle like the processes described earlier:
$e^{+} e^{-} \rightarrow \pi^{0} (\pi^{0'}) \gamma$ \cite{Arbuzov:2011fv}, $e^{+} e^{-} \rightarrow \pi^{0} \omega$ \cite{Arbuzov:2010xi},
$e^{+} e^{-} \rightarrow \pi^{0} \rho^{0}$ \cite{Ahmadov:2011ey}, $e^{+} e^{-} \rightarrow \left[\eta,\eta^{'},\eta(1295),\eta(1475)\right] \gamma$ \cite{Ahmadov:2013ksa}.

The obtained results are compared with the experimental data \cite{Aubert:2007ym, Ivanov:2016blh} and some theoretical results \cite{Chen:2013nna}.
Besides, a series of predictions is made.

\section{The Lagrangian of the extended NJL model for the mesons \\
$K^{\pm}, \eta, \eta', K^{*\pm}, \phi, \omega, \rho$}
In the extended NJL model, the quark-meson interaction Lagrangian for pseudoscalar $K^{\pm}, \eta, \eta'$,
vector $K^{*\pm}, \phi, \omega, \rho$ mesons in the ground and first radially excited states takes the form:

\begin{displaymath}
\Delta L_{int}(q,\bar{q},\eta,\eta'K,K^{*},\phi,\omega,\rho) = \bar{q}\left[i\gamma^{5}\sum_{j = \pm}\lambda_{j}(a_{K}K^{j} + b_{K}\hat{K}^{j})
\right.
\end{displaymath}
\begin{displaymath}
+ \frac{1}{2}\gamma^{\mu}\lambda_{\rho}(a_{\rho}\rho_{\mu} + b_{\rho}\hat{\rho}_{\mu})
+ \frac{1}{2}\gamma^{\mu}\lambda_{\omega}(a_{\omega}\omega_{\mu} + b_{\omega}\hat{\omega}_{\mu})
+ \frac{1}{2}\gamma^{\mu}\lambda_{\phi}(a_{\phi}\phi_{\mu} + b_{\phi}\hat{\phi}_{\mu})
\end{displaymath}
\begin{equation}
\left.+ \frac{1}{2}\gamma^{\mu}\sum_{j = \pm}\lambda_{j}(a_{K^{*}}K^{*j}_{\mu} + b_{K^{*}}\hat{K}^{*j}_{\mu})
+ i\gamma^{5} \sum_{j = u, s} \lambda_{j}
\sum_{\tilde{\eta} = \eta, \eta^{'}, \hat{\eta}, \hat{\eta}^{'}} A_{\tilde{\eta}}^{j} \tilde{\eta}\right]q,
\end{equation}
where $q$ and $\bar{q}$ are the u-, d- and s- constituent quark fields with masses $m_{u} = m_{d} = 280$ MeV,
$m_{s} = 420$ MeV \cite{Volkov:1999yi,Volkov:2001ns}, $\eta$, $\eta'$, $K^{\pm}$, $K^{*\pm}$, $\rho$, $\omega$ and $\phi$ are
the pseudoscalar and vector mesons, the excited states are marked with hat,

\begin{displaymath}
a_{a} = \frac{1}{\sin(2\theta_{a}^{0})}\left[g_{a}\sin(\theta_{a} + \theta_{a}^{0}) +
g_{a}^{'}f_{a}(\vec{k}^{2})\sin(\theta_{a} - \theta_{a}^{0})\right],
\end{displaymath}
\begin{equation}
\label{Coefficients}
b_{a} = \frac{-1}{\sin(2\theta_{a}^{0})}\left[g_{a}\cos(\theta_{a} + \theta_{a}^{0}) +
g_{a}^{'}f_{a}(\vec{k}^{2})\cos(\theta_{a} - \theta_{a}^{0})\right],
\end{equation}
\begin{displaymath}
A_{\tilde{\eta}}^{j} = g_{1}^{j}b_{\tilde{\eta}1}^{j} + g_{2}^{j}f_{j}(\vec{k}^{2})b_{\tilde{\eta}2}^{j},
\end{displaymath}
$f\left(\vec{k}^{2}\right) = 1 + d \vec{k}^{2}$ is the form factor for the description of the first radially excited states
\cite{Volkov:1996br, Volkov:1996fk}, $d$ is the slope parameter, $\theta_{a}$ and $\theta_{a}^{0}$ are
the mixing angles for the strange mesons in the ground and excited states

\begin{displaymath}
d_{uu} = -1.784 \textrm{GeV}^{-2}, \quad d_{us} = -1.761 \textrm{GeV}^{-2}, \quad d_{ss} = -1.737 \textrm{GeV}^{-2},
\end{displaymath}
\begin{equation}
\begin{array}{cccc}
\theta_{K} = 58.11^{\circ},    & \theta_{K^{*}} = 84.74^{\circ},    & \theta_{\rho} = \theta_{\omega} = 81.8^{\circ},        & \theta_{\phi} = 68.4^{\circ},\\
\theta_{K}^{0} = 55.52^{\circ},& \theta_{K^{*}}^{0} = 59.56^{\circ},& \theta_{\rho}^{0} = \theta_{\omega}^{0} = 61.5^{\circ},& \theta_{\phi}^{0} = 57.13^{\circ}.
\end{array}
\end{equation}

The insertion of the pseudoscalar isoscalar fields requires consideration of the mixing of the four
different states: $\eta, \eta'(958), \eta(1295), \eta(1475)$, which are marked as $\eta, \eta', \hat{\eta}, \hat{\eta}'$.
The last two ones are considered as the first radially excited states of the $\eta$ and $\eta'$ mesons;
$b_{\tilde{\eta}1}^{j}$ and $b_{\tilde{\eta}2}^{j}$ are the mixing coefficients shown in Table~\ref{TabCoeff} \cite{Volkov:1999yi}.

\begin{table}[h]
\caption{The mixing coefficients for the $\eta$-mesons.}
\label{TabCoeff}
\begin{center}
\begin{tabular}{ccccc}
                        & $\eta$ & $\hat{\eta}$ & $\eta'$ & $\hat{\eta}'$ \\
$b_{\tilde{\eta}1}^{u}$ & 0.71   & 0.62         & -0.32   & 0.56          \\
$b_{\tilde{\eta}2}^{u}$ & 0.11   & -0.87        & -0.48   & -0.54         \\
$b_{\tilde{\eta}1}^{s}$ & 0.62   & 0.19         & 0.56    & -0.67         \\
$b_{\tilde{\eta}2}^{s}$ & 0.06   & -0.66        & 0.30    & 0.82
\end{tabular}
\end{center}
\end{table}

These coefficients were successfully applied for description of a series of processes with the $\eta$-mesons
\cite{Volkov:1999yi, Arbuzov:2011rb}.

The matrices

\begin{displaymath}
\lambda_{\rho} = \left(\begin{array}{ccc}
1 & 0  & 0\\
0 & -1 & 0\\
0 & 0  & 0
\end{array}\right), \quad
\lambda_{\omega} = \left(\begin{array}{ccc}
1 & 0  & 0\\
0 & 1  & 0\\
0 & 0  & 0
\end{array}\right), \quad
\lambda_{\phi} = \sqrt{2} \left(\begin{array}{ccc}
0 & 0  & 0\\
0 & 0  & 0\\
0 & 0  & 1
\end{array}\right), \quad
\end{displaymath}

\begin{displaymath}
\lambda_{+} = \sqrt{2} \left(\begin{array}{ccc}
0 & 0 & 1\\
0 & 0 & 0\\
0 & 0 & 0
\end{array}\right), \quad
\lambda_{-} = \sqrt{2} \left(\begin{array}{ccc}
0 & 0 & 0\\
0 & 0 & 0\\
1 & 0 & 0
\end{array}\right),
\end{displaymath}

\begin{equation}
\lambda_{u} = \left(\begin{array}{ccc}
1 & 0 & 0\\
0 & 1 & 0\\
0 & 0 & 0
\end{array}\right), \quad
\lambda_{s} = -\sqrt{2} \left(\begin{array}{ccc}
0 & 0 & 0\\
0 & 0 & 0\\
0 & 0 & 1
\end{array}\right).
\end{equation}

The coupling constants:

\begin{displaymath}
g_{K} = \left(\frac{4}{Z_{K}}I_{2}(m_{u},m_{s})\right)^{-1/2} \approx 3.77,
\quad g_{K}^{'} = \left(4I_{2}^{f_{us}^{2}}(m_{u},m_{s})\right)^{-1/2} \approx 4.69,
\end{displaymath}
\begin{displaymath}
g_{K^{*}} = \left(\frac{2}{3}I_{2}(m_{u},m_{s})\right)^{-1/2} \approx 6.81,
\quad g_{K^{*}}^{'} = \left(\frac{2}{3}I_{2}^{f_{us}^{2}}(m_{u},m_{s})\right)^{1/2} \approx 11.49,
\end{displaymath}
\begin{displaymath}
g_{\rho} = g_{\omega} = \left(\frac{2}{3}I_{2}(m_{u},m_{u})\right)^{-1/2} \approx 6.14,
\quad g_{\rho}^{'} = g_{\omega}^{'} = \left(\frac{2}{3}I_{2}^{f_{uu}^{2}}(m_{u},m_{u})\right)^{1/2} \approx 9.87,
\end{displaymath}
\begin{displaymath}
g_{\phi} = \left(\frac{2}{3}I_{2}(m_{s},m_{s})\right)^{-1/2} \approx 7.5,
\quad g_{\phi}^{'} = \left(\frac{2}{3}I_{2}^{f_{ss}^{2}}(m_{s},m_{s})\right)^{1/2} \approx 13.19,
\end{displaymath}
\begin{displaymath}
g_{1}^{u} = \left(\frac{4}{Z_{\pi}}I_{2}(m_{u},m_{u})\right)^{-1/2} \approx 3.02,
\quad g_{2}^{u} = \left(4I_{2}^{f_{uu}^{2}}(m_{u},m_{u})\right)^{-1/2} \approx 4.03,
\end{displaymath}
\begin{equation}
\label{Constants}
g_{1}^{s} = \left(\frac{4}{Z_{s}}I_{2}(m_{s},m_{s})\right)^{-1/2} \approx 4.41,
\quad g_{2}^{s} = \left(4I_{2}^{f_{ss}^{2}}(m_{s},m_{s})\right)^{-1/2} \approx 5.39,
\end{equation}
where

\begin{displaymath}
Z_{\pi} = \left(1 - 6\frac{m^{2}_{u}}{M^{2}_{a_{1}}}\right)^{-1} \approx 1.45, \quad
Z_{s} = \left(1 - 6\frac{m^{2}_{s}}{M^{2}_{f_{1}}}\right)^{-1} \approx 2.09,
\end{displaymath}
\begin{equation}
Z_{K} = \left(1 - \frac{3}{2}\frac{(m_{u} + m_{s})^{2}}{M^{2}_{K_{1}}}\right)^{-1} \approx 1.83,
\end{equation}
$Z_{\pi}$ is the factor corresponding to the $\pi - a_{1}$ and $\eta - a_{1}$ transitions,
$Z_{K}$ is the factor corresponding to the $K - K_{1}$ transitions,
$Z_{s}$ is the factor corresponding to the $\eta - f_{1}$ transitions,
$M_{a_{1}} = 1230$ MeV, $M_{K_{1}} = 1272$ MeV, $M_{f_{1}} = 1426$ MeV \cite{Agashe:2014kda} are the masses
of the axial-vector $a_{1}$, $K_{1}$ and $f_{1}$ mesons, and the integral $I_{2}$ has the following form:

\begin{equation}
I_{2}^{f^{n}}(m_{1}, m_{2}) =
-i\frac{N_{c}}{(2\pi)^{4}}\int\frac{f^{n}(\vec{k}^{2})}{(m_{1}^{2} - k^2)(m_{2}^{2} - k^2)}\theta(\Lambda_{3}^{2} - \vec{k}^2)
\mathrm{d}^{4}k,
\end{equation}
$\Lambda_{3} = 1.03$ GeV is the cut-off parameter \cite{Volkov:1999yi}.

All these parameters were calculated earlier and are standard for the extended NJL model.

\section{The amplitude of the processes $e^{+}e^{-} \rightarrow K^{\pm} K^{*\mp}(892)$}
\begin{figure}[h]
\center{\includegraphics[scale = 0.6]{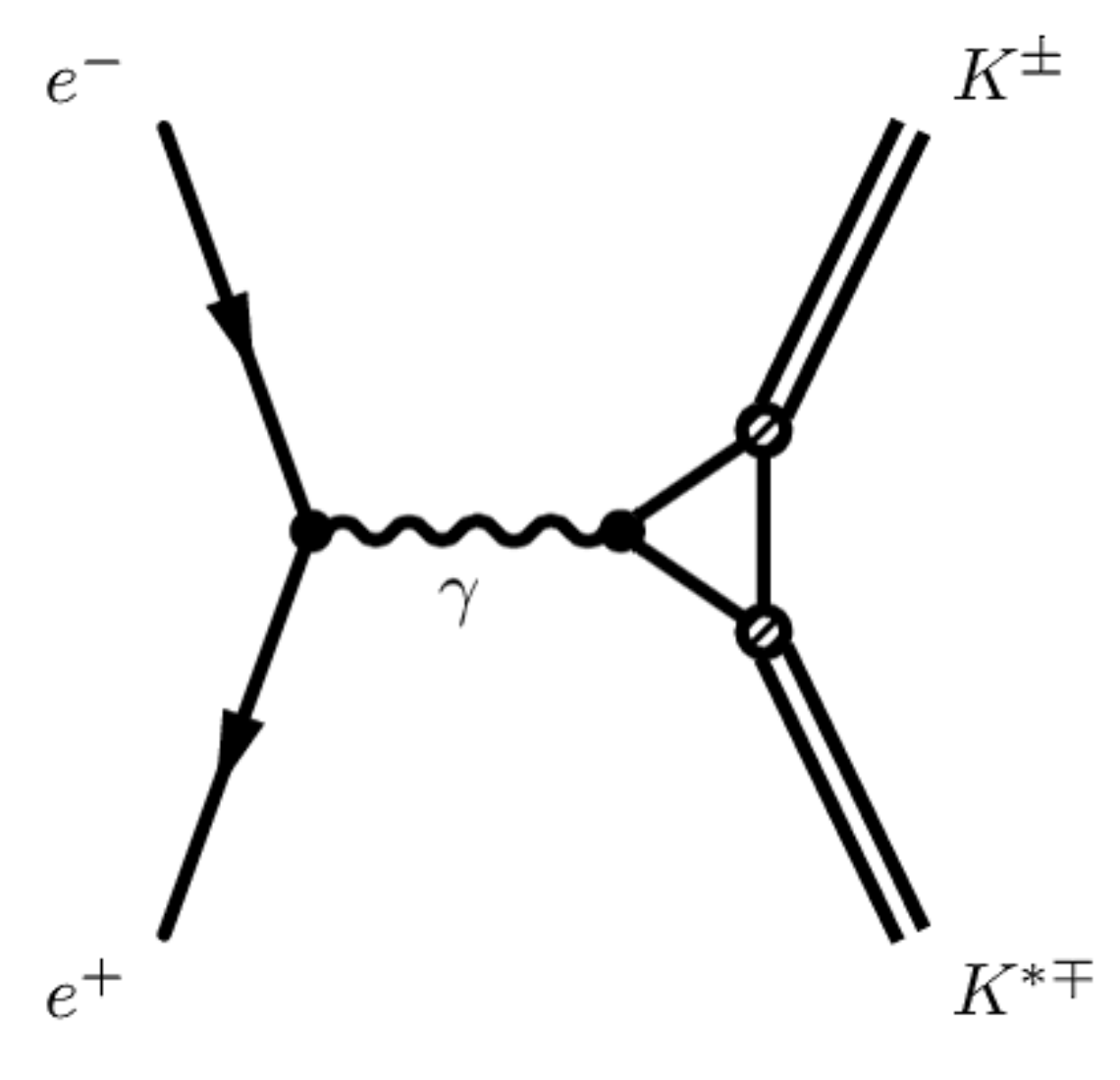}}
\caption{The processes $e^{+}e^{-} \rightarrow K^{\pm} K^{*\mp}(892)$ with an intermediate photon (contact diagram).}
\label{Contact1}
\end{figure}

\begin{figure}[h]
\center{\includegraphics[scale = 0.8]{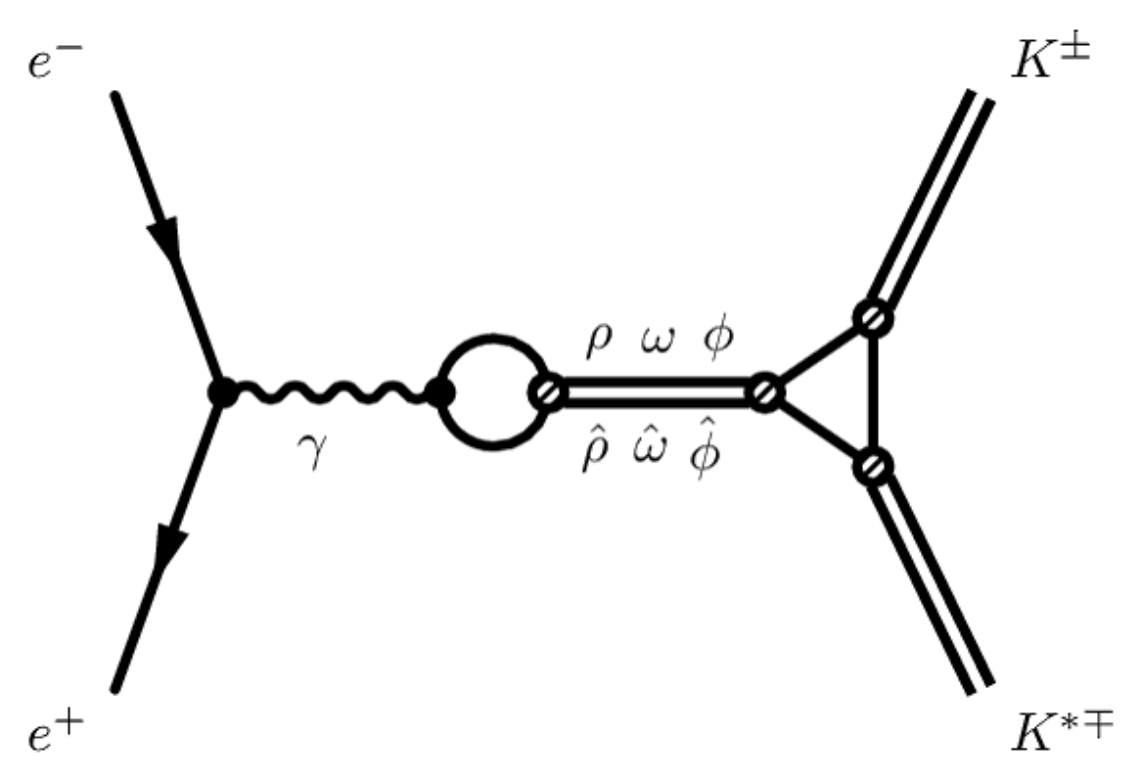}}
\caption{The processes $e^{+}e^{-} \rightarrow K^{\pm} K^{*\mp}(892)$ with the intermediate vector mesons.}
\label{Intermediate1}
\end{figure}

The diagrams of the processes $e^{+}e^{-} \rightarrow K^{\pm} K^{*\mp}(892)$ are shown in Figs.\ref{Contact1},\ref{Intermediate1}.


The appropriate amplitude takes the form:

\begin{equation}
T = \frac{8 \pi \alpha_{em}}{s} l^{\mu} \left\{B_{(\gamma)} + B_{(\rho + \hat{\rho})} + B_{(\omega + \hat{\omega})} +
B_{(\phi + \hat{\phi})}\right\}_{\mu\nu} e^{\nu\lambda\delta\sigma} p_{K\delta}p_{K^{*}\sigma} K^{*}_{\lambda},
\end{equation}
where $s = (p(e^{-}) + p(e^{+}))^2$, $l^{\mu} = \bar{e}\gamma^{\mu}e$ is the lepton current. The contribution of the
contact diagram is

\begin{equation}
B_{(\gamma)\mu\nu} = \frac{2}{3} \left[2m_{s}I^{\gamma K^{*}K}_{21}(m_{u}, m_{s}) - m_{u}I^{\gamma K^{*}K}_{12}(m_{u}, m_{s})\right] g_{\mu\nu},
\end{equation}
where $\gamma = 1$.

The contribution of the diagrams with the intermediate $\rho$-meson is

\begin{equation}
B_{(\rho + \hat{\rho})\mu\nu} = \frac{m_{s}}{g_{\rho}}
\left[C_{\rho}\frac{g_{\mu\nu}s - q_{\mu}q_{\nu}}{M^{2}_{\rho} - s - i\sqrt{s}\Gamma_{\rho}} I^{\rho K^{*}K}_{21}(m_{u}, m_{s})
+ e^{i\pi}C_{\hat{\rho}}\frac{g_{\mu\nu}s - q_{\mu}q_{\nu}}{M^{2}_{\hat{\rho}} - s - i\sqrt{s}\Gamma_{\hat{\rho}}} I^{\hat{\rho} K^{*}K}_{21}(m_{u}, m_{s})\right].
\end{equation}
The contribution of the diagrams with the intermediate $\omega$-meson is

\begin{equation}
B_{(\omega + \hat{\omega})\mu\nu} = \frac{m_{s}}{3g_{\omega}}
\left[C_{\omega}\frac{g_{\mu\nu}s - q_{\mu}q_{\nu}}{M^{2}_{\omega} - s - i\sqrt{s}\Gamma_{\omega}} I^{\omega K^{*}K}_{21}(m_{u}, m_{s})
+ e^{i\pi}C_{\hat{\omega}}\frac{g_{\mu\nu}s - q_{\mu}q_{\nu}}{M^{2}_{\hat{\omega}} - s - i\sqrt{s}\Gamma_{\hat{\omega}}} I^{\hat{\omega}K^{*}K}_{21}(m_{u}, m_{s})\right].
\end{equation}
The contribution of the diagrams with the intermediate $\phi$-meson is

\begin{equation}
B_{(\phi + \hat{\phi})\mu\nu} = -\frac{2m_{u}}{3g_{\phi}}
\left[C_{\phi}\frac{g_{\mu\nu}s - q_{\mu}q_{\nu}}{M^{2}_{\phi} - s - i\sqrt{s}\Gamma_{\phi}} I^{\phi K^{*}K}_{12}(m_{u}, m_{s})
+ e^{i\pi}C_{\hat{\phi}}\frac{g_{\mu\nu}s - q_{\mu}q_{\nu}}{M^{2}_{\hat{\phi}} - s - i\sqrt{s}\Gamma_{\hat{\phi}}} I^{\hat{\phi}K^{*}K}_{12}(m_{u}, m_{s})\right].
\end{equation}
$M_{\rho} = 775$ MeV, $M_{\hat{\rho}} = 1465$ MeV, $M_{\omega} = 783$ MeV, $M_{\hat{\omega}} = 1420$ MeV, $M_{\phi} = 1019$ MeV, $M_{\hat{\phi}} = 1680$ MeV,
$\Gamma_{\rho} = 149$ MeV, $\Gamma_{\hat{\rho}} = 400$ MeV, $\Gamma_{\omega} = 8$ MeV, $\Gamma_{\hat{\omega}} = 215$ MeV, $\Gamma_{\phi} = 4$ MeV,
$\Gamma_{\hat{\phi}} = 150$ MeV are the masses and the full widths of the intermediate vector mesons \cite{Agashe:2014kda}.

Unfortunately, the NJL model can not describe relative phase between ground and excited states. Thus, we should get phase from $e^{+} e^{-}$ annihilation
experiments \cite{Achasov:2006dv}. Similarly to our previous works \cite{Volkov:2016umo}, we use the factor $e^{i\pi}$ for the excited states.

The numerical coefficients $C_{a}$ are obtained from the quark loops in the transitions of the photon into the intermediate vector mesons:

\begin{equation}
C_{a} = \frac{1}{\sin\left(2\theta_{a}^{0}\right)}\left[\sin\left(\theta_{a} + \theta_{a}^{0}\right) +
R_{V}\sin\left(\theta_{a} - \theta_{a}^{0}\right)\right],
\end{equation}

\begin{displaymath}
R_{V} = \frac{I_{2}^{f}(m_{1},m_{2})}{\sqrt{I_{2}(m_{1},m_{2})I_{2}^{f^{2}}(m_{1},m_{2})}},
\end{displaymath}
where $m_{1}$ and $m_{2}$ are the masses of the u-quarks or the s-quarks depending on the quark structure of the intermediate vector meson.
The integrals

\begin{equation}
I^{abc}_{mn}(m_{u}, m_{s}) = -i\frac{N_{c}}{(2\pi)^{4}}\int\frac{a(\vec{k}^{2})b(\vec{k}^{2})c(\vec{k}^{2})}{(m_{u}^{2} - k^2)^{m}(m_{s}^{2} - k^2)^{n}}
\theta(\Lambda_{3}^{2} - \vec{k}^2) \mathrm{d}^{4}k,
\end{equation}
are obtained from the quark triangles, $a(\vec{k}^{2})$, $b(\vec{k}^{2})$ and $c(\vec{k}^{2})$ are the coefficients from the Lagrangian defined in (\ref{Coefficients}).

\section{The amplitude of the process $e^{+}e^{-} \rightarrow \eta \phi(1020)$ in the extended NJL model}
\begin{figure}[h]
\center{\includegraphics[scale = 0.6]{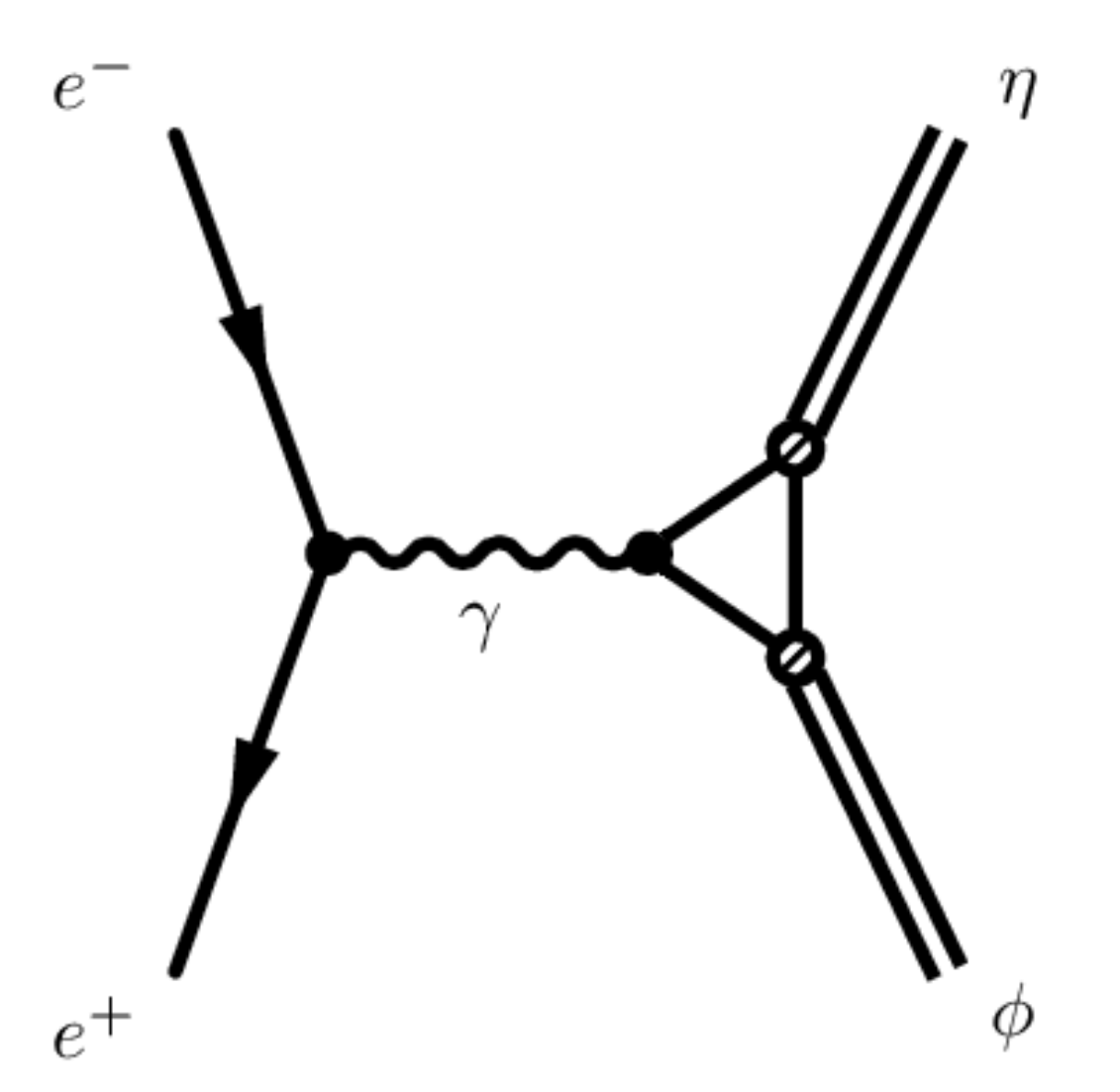}}
\caption{The process $e^{+}e^{-} \rightarrow \eta \phi(1020)$ with an intermediate photon (contact diagram).}
\label{Contact2}
\end{figure}

\begin{figure}[h]
\center{\includegraphics[scale = 0.8]{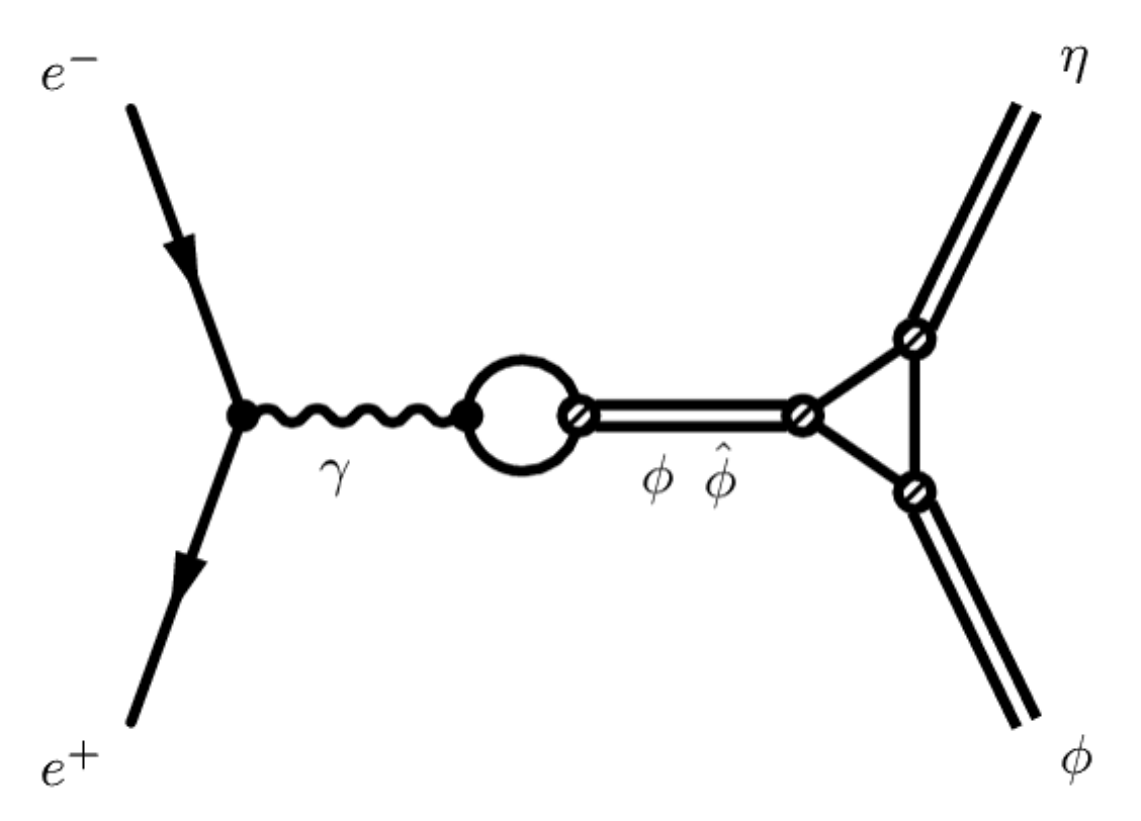}}
\caption{The process $e^{+}e^{-} \rightarrow \eta \phi(1020)$ with the intermediate vector mesons.}
\label{Intermediate2}
\end{figure}

The diagrams of the process $e^{+}e^{-} \rightarrow \eta \phi(1020)$ are shown in Figs.\ref{Contact2},\ref{Intermediate2}.


The appropriate amplitude takes the form:

\begin{displaymath}
T = \frac{4 \pi \alpha_{em}}{s} l^{\mu} \frac{8m_{s}}{3s} \left\{I^{\gamma\phi\eta}_{03}(m_{u}, m_{s})g_{\mu\nu}
+ \frac{C_{\phi}}{g_{\phi}} I^{\phi\phi\eta}_{03}(m_{u}, m_{s})
\frac{g_{\mu\nu}s - q_{\mu}q_{\nu}}{M^{2}_{\phi} - s - i\sqrt{s}\Gamma_{\phi}}\right.
\end{displaymath}
\begin{equation}
\left. + e^{i\pi}\frac{C_{\hat{\phi}}}{g_{\phi}} I^{\hat{\phi}\phi\eta}_{03}(m_{u}, m_{s})
\frac{g_{\mu\nu}s - q_{\mu}q_{\nu}}{M^{2}_{\hat{\phi}} - s - i\sqrt{s}\Gamma_{\hat{\phi}}}\right\}
e^{\nu\lambda\delta\sigma} p_{\eta\delta}p_{\phi\sigma} \phi_{\lambda}.
\end{equation}

\section{Numerical estimations}
\begin{figure}[h]
\center{\includegraphics[scale = 0.7]{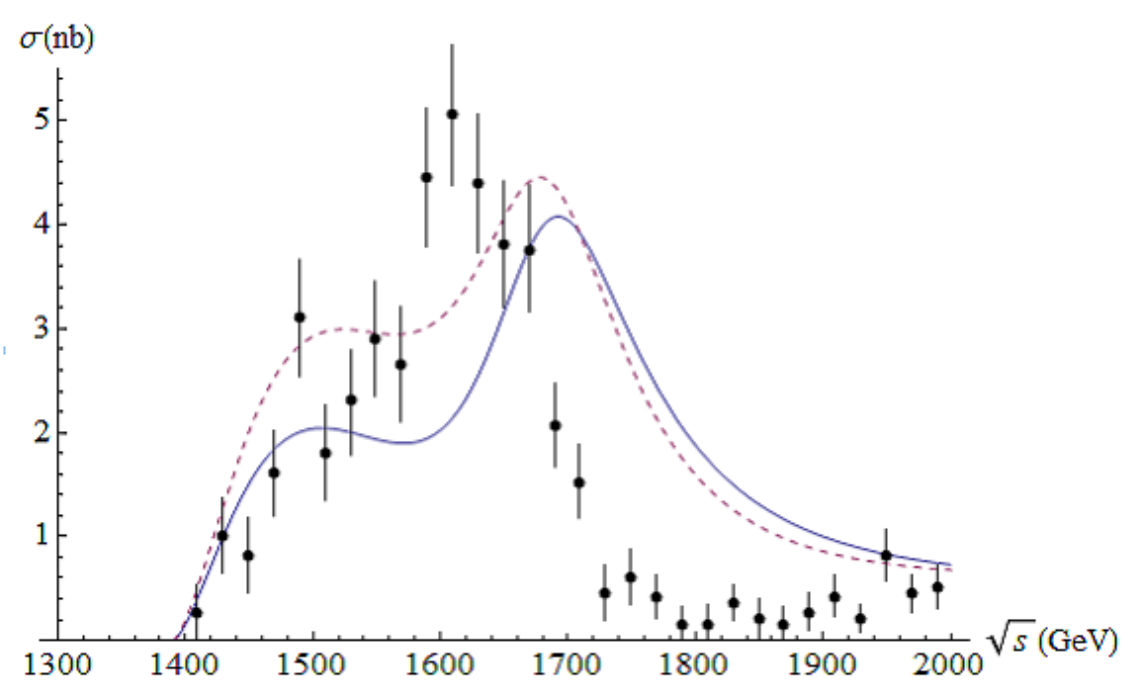}}
\caption{The cross section of the processes $e^{+}e^{-} \rightarrow K^{\pm} K^{*\mp}(892)$.}
\label{CrossSection1}
\end{figure}

The cross section of the processes $e^{+}e^{-} \rightarrow K^{\pm} K^{*\mp}(892)$ is shown in Fig.\ref{CrossSection1}.
The solid line corresponds to our theoretical cross section. The points correspond to the experimental values \cite{Aubert:2007ym}.
It is interesting to note, that the minimal value of the full width of $\rho(1450)$ ($\Gamma_{\hat{\rho}} = 340$ MeV) provides
a slight left shift and increase the theoretical peak. The appropriate theoretical cross section is shown by dashed line.

\begin{figure}[h]
\center{\includegraphics[scale = 0.7]{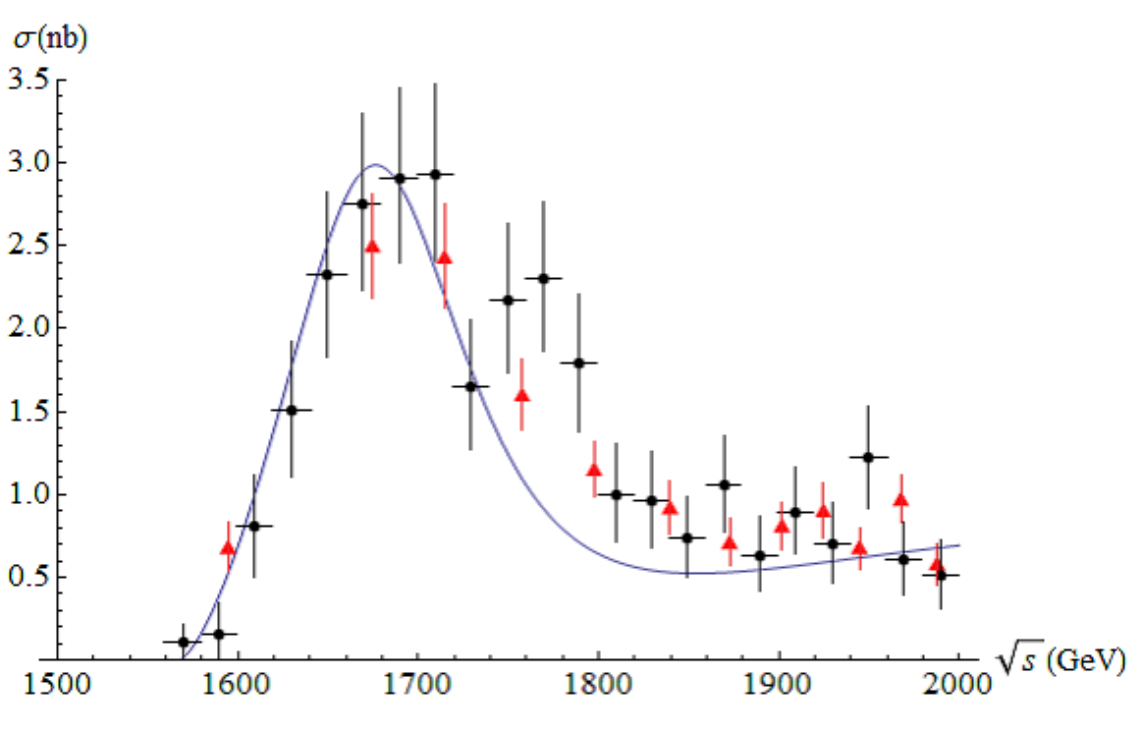}}
\caption{The cross section of the process $e^{+}e^{-} \rightarrow \eta \phi(1020)$.}
\label{CrossSection2}
\end{figure}

The cross section of the process $e^{+}e^{-} \rightarrow \phi(1020) \eta$ is shown in Fig.\ref{CrossSection2}.
The solid line corresponds to our theoretical cross section. The triangle points correspond to the experimental values from
CMD-3 \cite{Ivanov:2016blh}, the round points correspond to the experimental values from the BaBar Collaboration \cite{Aubert:2007ym}.

The processes with excited mesons in the final states and with the $\eta^{'}(958)$ meson were calculated similarly.

\begin{figure}[h!]
\center{\includegraphics[scale = 0.7]{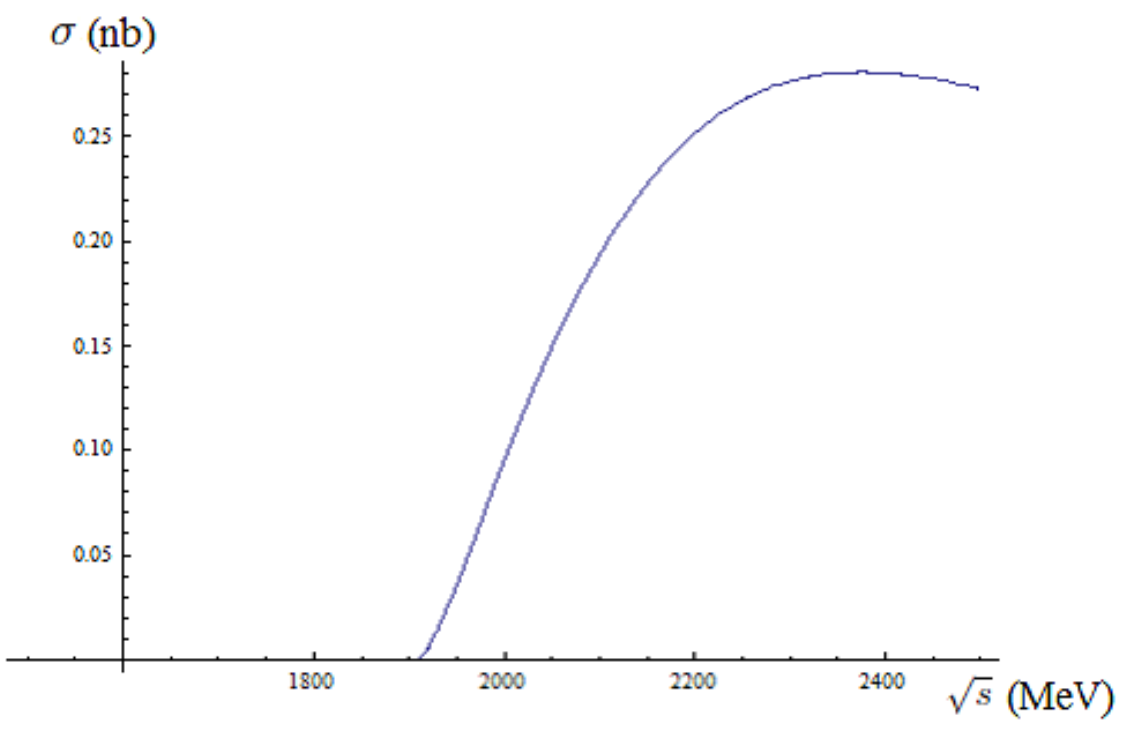}}
\caption{The cross section of the processes $e^{+}e^{-} \rightarrow K^{\pm} K^{*\mp}(1410)$.}
\label{CrossSection3}
\end{figure}

\begin{figure}[h!]
\center{\includegraphics[scale = 0.7]{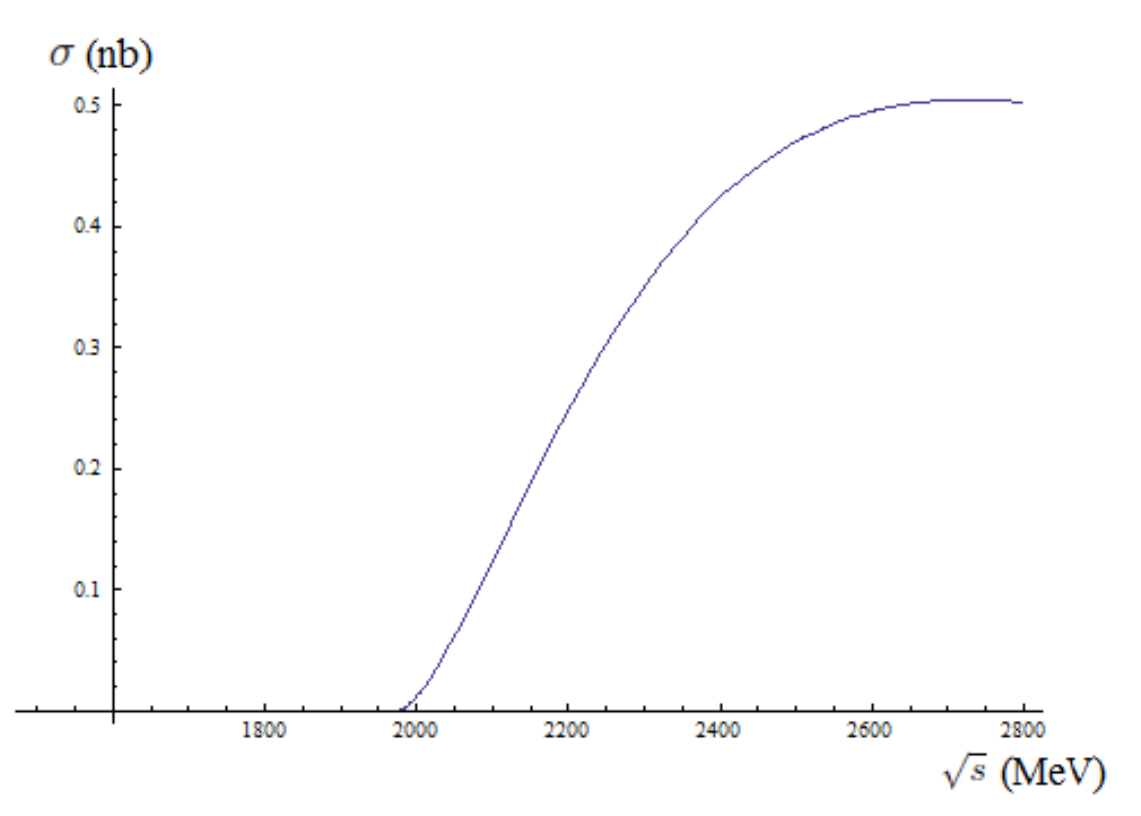}}
\caption{The cross section of the process $e^{+}e^{-} \rightarrow \eta^{'} \phi(1020)$.}
\label{CrossSection4}
\end{figure}

\begin{figure}[h!]
\center{\includegraphics[scale = 0.7]{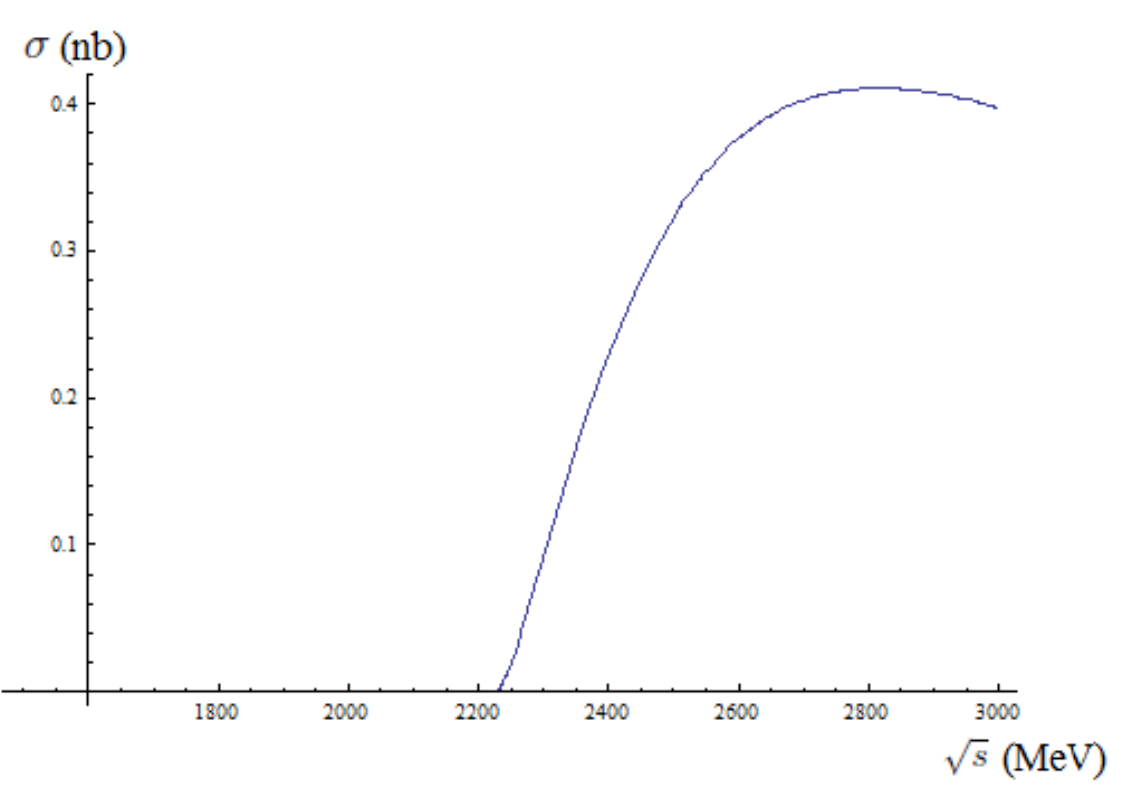}}
\caption{The cross section of the process $e^{+}e^{-} \rightarrow \eta \phi(1680)$.}
\label{CrossSection5}
\end{figure}

The cross section of the processes $e^{+}e^{-} \rightarrow K^{\pm} K^{*\mp}(1410)$, $e^{+}e^{-} \rightarrow \eta^{'} \phi(1020)$
and $e^{+}e^{-} \rightarrow \eta \phi(1680)$ are shown in Fig.\ref{CrossSection3},\ref{CrossSection4},\ref{CrossSection5}.

The last three predictions are in the energy range where the NJL model can provide only qualitative estimations.

\section{Conclusion}
As already mentioned in Introduction, the NJL model does not require any additional fitting parameters for description
of the experimental data. Indeed, it includes a small amount of the model parameters to describe the meson interactions:
quark masses, the cut-off parameter and the 't Hooft parameter.
The spontaneous breaking of chiral symmetry of strong interactions have allowed us to describe different
processes of meson interactions at low energies \cite{Volkov:2016umo, Volkov:1993jw, Ebert:1994mf, Volkov:2006vq, Volkov:1999yi}.
Therefore, describing a wide range of processes, we can not provide high precision for each of them. However, we obtain
satisfying agreement with the experimental data.

The main peak of cross section of the processes $e^{+}e^{-} \rightarrow K^{\pm} K^{*\mp}(892)$ calculated in this work is shifted
to the right with respect to experimental points.
However, the minimal value of the full width of $\rho(1450)$ provides a slight left shift of the theoretical peak.
The theoretical description of this reaction was successfully made in \cite{Chen:2013nna} within the framework of
chiral effective field theory, but the fitting parameters were used there and the first peak was not described.

The cross section of the process $e^{+}e^{-} \rightarrow \eta \phi(1020)$ obtained in the present paper, include one
resonance which is consistent with the experimental data from BaBar and CMD-3. However the BaBar experimental results
include the second peak which is absent in the results from CMD-3. Thus the behavior of our theoretical cross section
is in better agreement with the CMD-3 experimental data.

The reactions considered in the present work may be useful for the description of the processes $e^{+}e^{-} \rightarrow K K (\eta, \eta^{'}(958), \pi)$.
The qualitative estimations of the cross sections of the processes with the first radially-excited mesons
and $\eta^{'}$ meson in the final states are made.

\section*{Acknowledgments}
We are grateful to A. B. Arbuzov and O. V. Teryaev for useful discussions; this work is supported by the RFBR grant, 14-01-00647.

\end{document}